\documentclass[a4paper]{article}
\usepackage[utf8]{inputenc}
\usepackage[UKenglish]{babel}
\usepackage{graphicx}
\usepackage{units}
\usepackage{opex3}
\usepackage{subfig}

\begin{document}

\addtolength{\abovecaptionskip}{-2mm}

\title{Demonstration of UV-written waveguides, Bragg gratings and cavities at \unit[780]{nm}, and an original experimental measurement of group delay}
\vskip4pc

\tableofcontents
\clearpage

\title{Demonstration of UV-written waveguides, Bragg gratings and cavities at \unit[780]{nm}, and an original experimental measurement of group delay}
\author{G. Lepert$^{1*}$, M. Trupke$^{1,2}$, E. A. Hinds$^1$, H. Rogers$^3$, J. C. Gates$^3$ and P. G. R. Smith$^3$}
\address{$^1$Centre for Cold Matter, Blackett Laboratory, Imperial College, London SW7 2AZ, UK

$^2$Current address: Vienna Center for Quantum Science and Technology, Atominstitut, TU-Wien, 1020 Vienna, Austria

$^3$Optoelectronics Research Centre, University of Southampton, Southampton, SO17 1BJ, UK
}
\email{guillaume.lepert07@imperial.ac.uk}

	\begin{abstract}
We present direct UV-written waveguides and Bragg gratings operating at \unit[780]{nm}. By combining two gratings into a Fabry-Perot cavity we have devised and implemented a novel and practical method of measuring the group delay of Bragg gratings.
	\end{abstract}

\bibliographystyle{osajnl}

\section{Introduction: from telecom fibre Bragg gratings to near-visible planar Bragg gratings}
	Fibre Bragg gratings (FBGs) have found many applications in optoelectronics, from distributed feedback lasers to temperature and stress sensors. Closely linked with the telecom industry, their operation has been overwhelmingly concentrated in the near infrared part of the spectrum (0.98 to 1.6 microns). But for applications in biochemistry or quantum optics, where such devices are also of considerable interest, Bragg gratings have to be integrated into glass or silicon-based planar devices and their operating wavelength has to be significantly shifted towards the visible spectrum.

``Lab-on-a-chip" devices, typically built around centimetre-scale glass or silicon-based planar devices, open a route to miniaturising the analysis of chemical and biological reactions thereby reducing the quantity of reactants required and the process time. One of the key requirements of this technology is an accurate method of monitoring environmental conditions and/or reaction progress. Fibre Bragg gratings can provide such monitoring by allowing part of the optical mode to interact with an analyte to measure small refractive index changes. It is therefore desirable to build Bragg gratings directly into the device. This can easily be achieved by direct UV-writing, as has been demonstrated in \cite{Sparrow2005} at \unit[1550]{nm}, where removing the cladding provided access for a liquid analyte to interact with the optical mode of the waveguide. Such devices can display refractive index sensitivities better than $10^{-5}$ at this wavelength. However this sensitivity can be greatly compromised for water-based analytes due to the strong absorption of OH bonds, in excess of \unit[30]{dB/cm} above \unit[1400]{nm} \cite{Hale1973}, which reduces the finesse of grating and etalon-based sensors.

Integrated photonic circuits will also play a very important part in future developments in quantum optics, in particular cavity quantum electrodynamics (QED), which exploits the interaction of light and matter (usually atoms) in the presence of a light-confining structure. For example, it is possible to enhance or suppress spontaneous emission from quantum emitters (Purcell effect) with the assistance of an optical resonator. Cavity QED  has witnessed extraordinary progresses in the past decade, and much activity in this field is now directed to moving beyond single cavity experiments and towards networks of cavities that can process quantum information e.g.~\cite{Lepert2011}. This requires a switch from bulk to integrated optics. UV-written waveguides are particularly well suited to such a research environment as they offer high flexibility in the design of photonic circuits together with fast and low-cost prototyping. However, existing work in UV-written waveguides has been restricted to the telecom band near $\unit[1.5]{ \mu m }$, where there is unfortunately no suitable atomic transition; a widely used wavelength is that of rubidium at \unit[780]{nm}.

The work reported in this article demonstrates that direct UV writing can be used to fabricate waveguides, Bragg gratings and etalons in the near-visible region of \unit[780]{nm} for application in sensing water based analytes or for use in quantum information processing. These are considerably more challenging to produce than at \unit[1500]{nm} because they require a reduction in both the spot size of the writing beam and in the grating period which places higher demands on the precision and stability of the writing system.

We focus our attention on Fabry-Perot etalons consisting of two nominally identical Bragg gratings. As sensors, these offer higher sensitivity than single gratings by virtue of the narrow resonance linewidth. As QED cavities they offer a new platform for processing quantum information through optical networks. An interesting feature of such Bragg cavities is the ability to determine all the properties both of a single grating (reflectivity, spectral profile, group delay) and of the waveguide itself (losses) by performing a small number of simple and straightforward measurements on the cavity. In particular, the group delay -- often a key property in the design of optical filters and delay lines -- is usually measured by non-trivial means such as white light interferometry or pulse delay measurement. But as we will demonstrate, it is straightforward to measure the group delay via its effect on the free spectral range of a Bragg cavity.

\section{UV-writing of optical waveguides and Bragg gratings at 780 nm}

\begin{figure}[t]
	\addtolength{\abovecaptionskip}{+2mm}
	\centering
	\includegraphics[width=\textwidth]{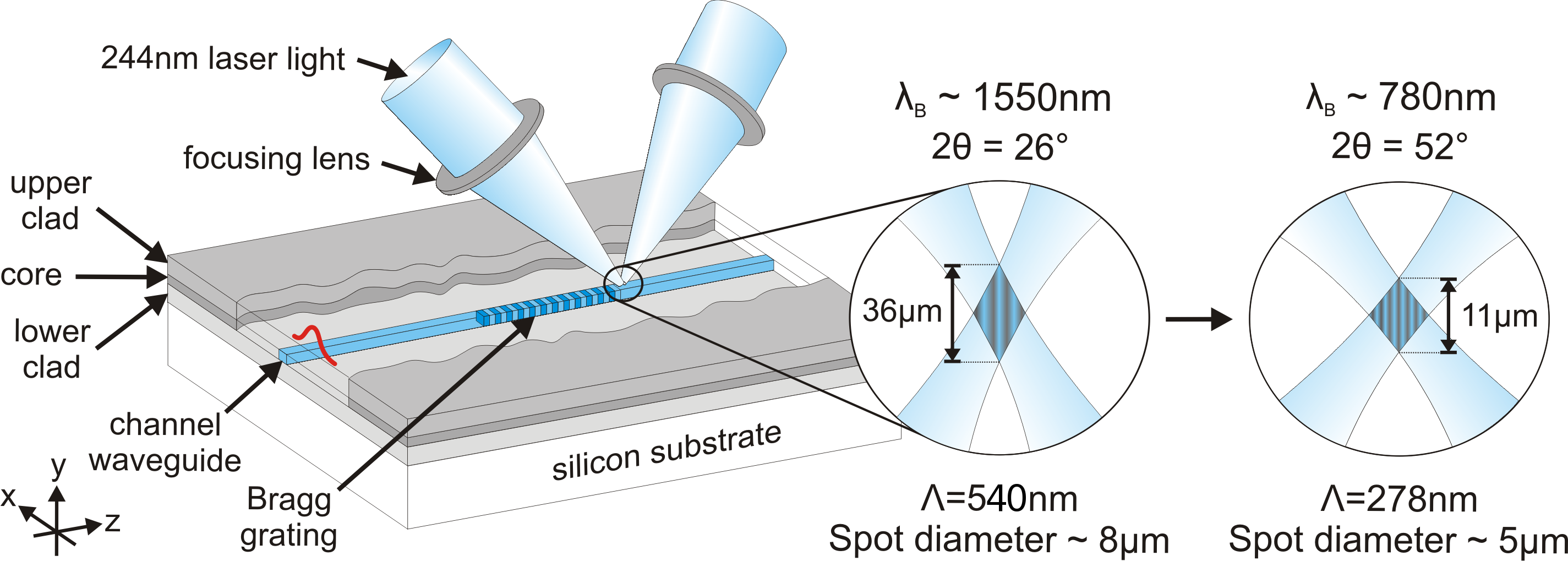}
	\caption{Direct UV writing scheme using 244 nm light. The Bragg wavelength $\Lambda$ of the gratings is determined by the angle $2\theta$ between the two interfering beams according to $\Lambda=244$\,nm$/(2 \sin \theta)$).}
	\label{fig:Kundys2009_Schematic}
\end{figure}

	Direct UV writing \cite{Svalgaard1994} relies on the photosensitivity to UV light of germanium-doped fused silica. A waveguide chip consists of a silicon substrate on top of which three silica layers are deposited. A first, thick ($\approx \unit[16]{ \mu m }$) layer of thermal oxide, grown by oxidation in an oxygen-rich furnace for one month, forms the underclad and is vital to reduce wafer bow which is a result of the different expansion coefficients of silica and silicon. The photosensitive middle layer, $\unit[5.6]{\mu m}$ deposited by flame hydrolysis, is doped with germanium, which provides photosensitivity, and with boron in order to match the refractive indices of the adjacent layers. This is covered by a $\unit[17]{\mu m}$-thick silica layer also deposited by flame hydrolysis. A \unit[244]{nm} CW laser (power approx. \unit[40]{mW}) is split at a 50:50 beam splitter and the two beams are focused in the core layer, where they produce an interference pattern with a period $\Lambda$ defined by the angle $2\theta$ between the two beams (Figure \ref{fig:Kundys2009_Schematic}). High precision air bearing stages with nanometre resolution move the substrate relative to the beams to produce the desired waveguide circuit. The waveguide index contrast can be modified by controlling the translation speed, as a lower speed results in higher fluence, hence a larger index change.

	Translation of the sample under the CW laser spot averages out the interference pattern, resulting in a uniform waveguide. Bragg gratings are written simultaneously, by switching the UV laser on and off with an acousto-optic modulator (AOM), the switching period being close to the time it takes for the sample to move through one period of the interference pattern. This stroboscopic method produces a periodic modulation of the refractive index, i.e. a Bragg grating. Within the bandwidth allowed by the UV fringe pattern, the period of this grating can be fine-tuned by adjusting the period of the laser modulation. The on-off duty cycle controls the index contrast, making it easy to apodise the grating, suppressing the side lobes which otherwise appear in the reflection spectrum. The spot diameter of the writing laser is much larger than the grating period ($\unit[5]{\mu m}$ \textit{vs.} \unit[263]{nm} for a Bragg wavelength of \unit[780]{nm}), so many laser modulation cycles contribute to any given Bragg layer. This averages out any jitter of the AOM timing and ensures that the grating period is very accurately controlled, a key requirement in view of the large number of layers involved (4900 grating planes for a \unit[1.3]{mm} long grating). Resonators are formed by writing two separated Bragg mirrors into a waveguide.

	When switching from \unit[1550]{nm} waveguides and gratings to \unit[780]{nm}, two major changes have to be made: the shorter Bragg wavelength requires a larger angle $2\theta$ between the two interfering beams, as illustrated in Figure \ref{fig:Kundys2009_Schematic}; and the core needs to be smaller to achieve single mode waveguides, which requires a smaller spot size. Both requirements conspire to reduce significantly the tolerance to alignment errors. An additional source of error comes from bow in the planar sample which causes a variation of the overlap between the inference pattern and the core layer, resulting in variation of the index contrast.

\begin{figure}[t]
	\centering
	\includegraphics[width=\textwidth]{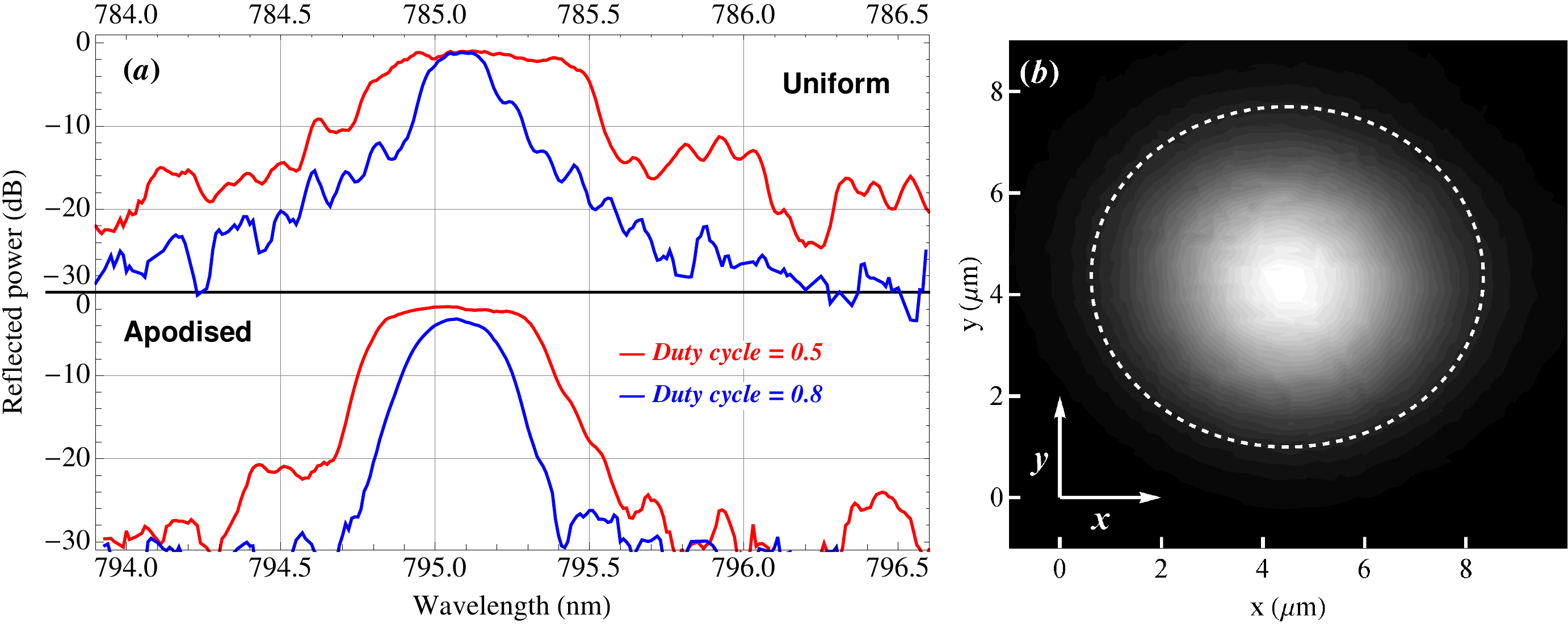}
	\caption{
	{\bf (a)} Experimental reflection spectra of uniform (top) and apodised (bottom) gratings, written with duty cycles (fraction of time on) of 0.5 (red) or 0.8 (blue). Apodisation removes the side lobes and increases the bandwidth. Higher duty cycle results in weaker gratings. The two gratings were written on a single waveguide, their design wavelengths being \unit[10]{nm} apart.
	{\bf (b)} Mode profile of a UV-written waveguide, measured by scanning a single-mode fibre in front of the waveguide. The dashed white line is a fit to the $1/e^2$ contour. After de-convolving the $\unit[5.0]{ \mu m }$ fibre mode, this gives a waveguide mode diameter of $\unit[5.8]{ \mu m }$ by $\unit[4.5]{ \mu m }$.}
	\label{fig:ExpExamples}
\end{figure}

	Examples of waveguides and gratings written in this way are shown in Fig.~\ref{fig:ExpExamples}(a). The reflection spectra of both uniform and apodised gratings were measured using a fibre-coupled SLED source (Exalos Ltd), a \unit[780]{nm} \unit[3]{dB coupler} and an optical spectrum analyser (Ando AQ 6317B). The spectra have been normalised to remove the wavelength dependence of the fibre coupler and optical source. In Fig.~\ref{fig:ExpExamples}(b) we present a measurement of the mode profile of a waveguide, obtained by coupling light into and out of the waveguide with two single-mode fibres and scanning one of the fibres with a piezo-actuated flexure stage (Thorlabs NanoMax 300). The mode size ($1/e^2$ intensity diameter), after deconvolving the measured fibre mode size ($\unit[5.0]{ \mu m }$), is found to be $\unit[5.8]{ \mu m }$ along the $x$-axis and $\unit[4.5]{ \mu m }$ along $y$ (axes are defined in Fig.~\ref{fig:Kundys2009_Schematic}). Such a quasi-circular mode can in principle couple to the fibre with an efficiency in excess of 95\%.

	With the exception of the waveguides used to measure the spectra and mode profile we just described, the present experiment used a single chip, containing a total of 30 waveguide resonators with cavity lengths between 4 and \unit[16]{mm} and mirror lengths between 1.3 and \unit[1.8]{mm}.

\section{Bragg grating and Bragg cavity theory: group delay and cavity length}

	The behaviour of light within Bragg gratings is usually described by means of coupled mode theory (see \cite{Erdogan1997} for a comprehensive review). Analytical solutions to the coupled mode equations exist in the case of uniform, sinusoidal gratings, while for more complex structures the equations can easily be integrated numerically, requiring only the knowledge of the envelope of the index profile along the grating. In this way, it is possible to determine the amplitude and phase of the field everywhere, not only for a single grating, but also for any combination of gratings. Moreover, losses can be included straightforwardly in the coupled mode equations through a complex refractive index.

\begin{figure}[bt]
	\centering
	\includegraphics[width=\textwidth]{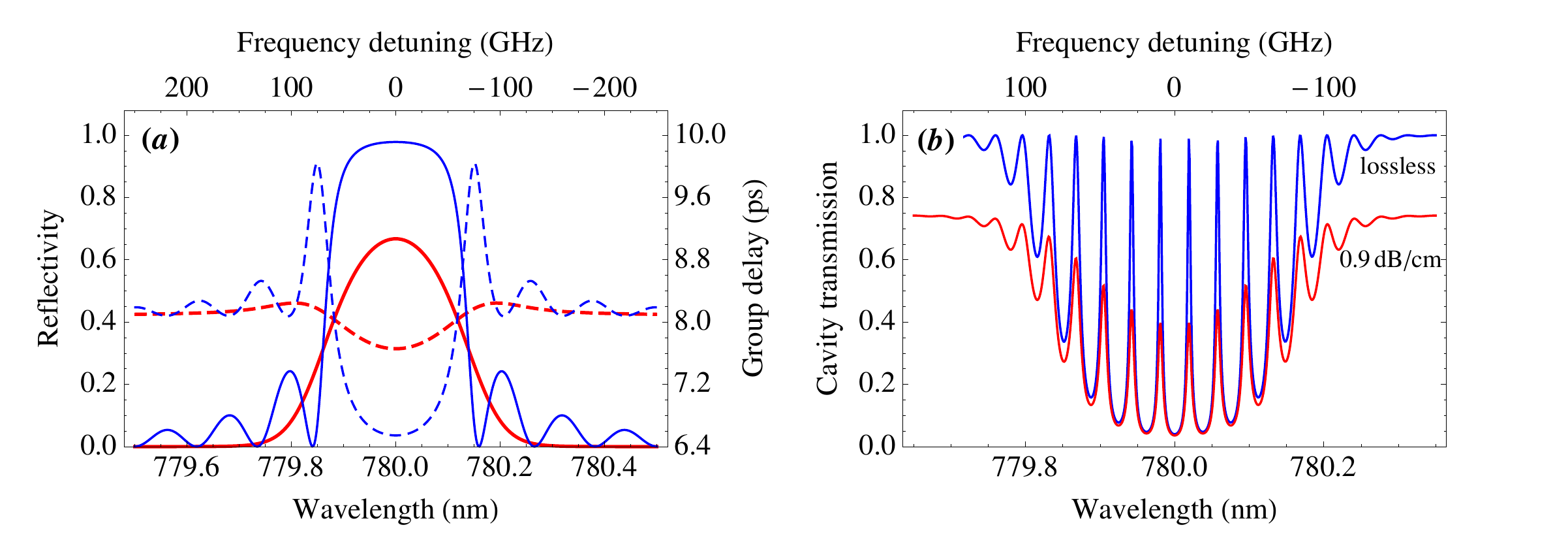}
	\caption{{\bf (a)} Reflectivity (solid lines) and group delay (dashed) of uniform (blue) and Gaussian-apodised (red) Bragg gratings, plotted as a function of wavelength $\lambda$ and frequency detuning from the peak $\Delta\nu$, calculated from coupled mode theory. Both gratings are $\unit[1.7]{mm}$ or about 6350 periods long, and the index contrast is set to $\Delta n = 3.8\times 10^{-4}$. The peak power reflection coefficients are 0.98 and 0.67 respectively. {\bf (b)} Calculated transmission spectrum of a cavity made up of two such Gaussian-apodised gratings separated by $\unit[4]{mm}$. The blue curves shows the lossless case whereas the red curve includes realistic propagation losses of \unit[0.9]{dB/cm}.}
	\label{fig:FSR&GroupDelay}
\end{figure}
	
	Figure \ref{fig:FSR&GroupDelay}(a) shows the theoretical reflection spectrum of a \unit[1.7]{mm}-long grating with index contrast $\Delta n = 3.8\times 10^{-4}$, in the uniform (blue lines) and Gaussian-apodised (red lines) cases, together with the group delay (dashed lines). In the apodised case, $\Delta n$ is the peak index contrast. The group delay exhibits a minimum at the Bragg wavelength as the higher reflectivity causes the reflected light to travel a shorter distance within the grating. The side lobes have virtually disappeared in the apodised version. Figure \ref{fig:FSR&GroupDelay}(b) shows in red the transmission spectrum of a cavity made of two such mirrors, separated by a \unit[4]{mm} gap, incorporating realistic propagation losses of \unit[0.9]{dB/cm} (i.e. about 25\% power loss per pass). Compared with the lossless cavity (in blue), the spectrum exhibits the usual Fabry-Perot resonance fringes, but we also notice that the transmission on resonance does not reach unity (after normalisation to the out of band transmission). This is the combined effect of losing more photons when there are multiple bounces in the cavity and of the effective mirror mismatch, which allows nonzero reflection at resonance.

	The connection between the single grating group delay and the cavity free spectral range (FSR) is best demonstrated using Figure \ref{fig:FSR&GroupDelay}(b) as an example, where the FSR is \unit[19.01]{GHz}. Since FSR$=c/(2L_{\mathrm{cav}})$, the optical cavity length is $L_{\mathrm{cav}} = \unit[7.89]{mm}$. Taking the refractive index of the guide as 1.46, the physical cavity length becomes $L_{\mathrm{eff}} = \unit[5.40]{mm}$. This is longer than the distance $L=\unit[4]{mm}$ between the innermost planes of the two gratings because the light penetrates inside the gratings; the penetration length $L_g=(L_{\mathrm{eff}}-L)/2$ \cite{Barmenkov2006} is directly linked to the grating group delay $\tau_g = d\phi/d\omega$ \cite{Erdogan1997} by $\tau_g=2 L_g n/c$. Indeed, the group delay calculated in this way ($\unit[6.84 \pm 0.04]{ps}$) matches exactly the coupled mode theory calculation of Fig.~\ref{fig:FSR&GroupDelay}(a).

	Thus grating cavities offer straightforward access to the group delay of gratings, a quantity whose determination otherwise requires complex white light interferometry techniques or pulse delay measurements. Providing the cavity is long enough, the group delay can even be spectrally resolved by observing the evolution of the FSR across the grating bandwidth, as we demonstrate in Section \ref{Lg_profile}. It should also be pointed out that not only the group delay but virtually all the properties of the grating and waveguide alike can be experimentally retrieved from a single cavity spectrum like the one presented above, as there is a direct correspondence between the finesse and transmission on the one hand and reflectivity and losses on the other.

\section{Group delay measurement: experimental results}

\subsection{Apparatus}
\begin{figure}[t]
	\centering
	\includegraphics[width=\textwidth]{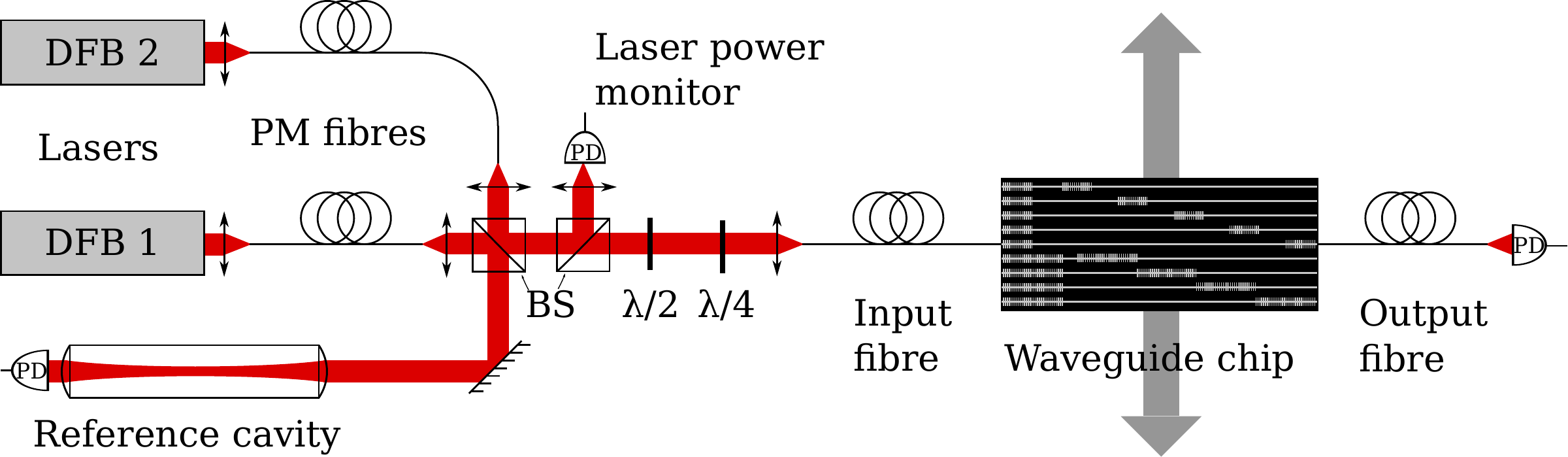}
	\caption{Schematic representation of the experimental setup. On the chip, a typical set of cavities with different grating and cavity lengths.}
	\label{fig:ExperimentalSetup}
\end{figure}

The experimental setup is illustrated in Figure \ref{fig:ExperimentalSetup}. Light from two \unit[780]{nm} distributed feedback (DFB) laser diodes (Eagleyard Photonics) is coupled to polarisation-maintaining fibres, and then combined at a beam splitter so that either laser can be used to probe the waveguides. Typically one laser is tuned far from the Bragg frequency and is used to measure the transmission through the waveguides in the ``absence" of gratings. The second laser is tuned over \unit[100]{GHz} by scanning the drive current of the laser diode, and beyond that by changing the temperature of the diode. Half- and quarter-wave plates adjust the polarisation before the light is coupled into the single-mode input fibre. The exit of this fibre is cleaved and aligned with a waveguide using a precision three-axis flexure stage. A second such stage aligns the output fibre, which is then fed to a photodiode. Undesirable etalon effects between the fibres and the waveguide chip are avoided by using an index-matching oil.

A second beam splitter sends a fraction of the light to a photodiode used as a normalisation to compensate for variations of the laser power during a frequency scan. Finally, part of the scanning laser is diverted to a reference cavity consisting of a \unit[5]{cm} long block of glass, with two reflection-coated lenses glued to the ends of the block. The assembly sits in a heavy aluminium housing that provides thermal inertia. A cavity finesse of 30 is measured, in agreement with the nominal mirror reflectivity of 90\%. The free spectral range is measured to be $\mathrm{FSR_{ref}}=\unit[1.905(5)]{GHz}$. During a laser scan we monitor the fringes of light transmitted by this cavity to obtain a calibrated relative frequency scale.

\subsection{Measurement method}
\begin{figure}[b]
	\centering
	\begin{minipage}{0.6\textwidth}
		\includegraphics[width=\textwidth]{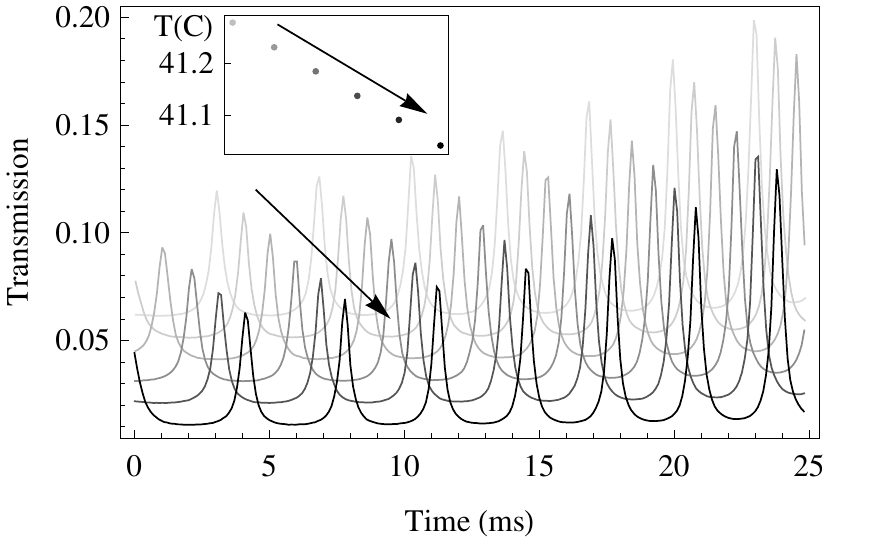}
	\end{minipage}
	\begin{minipage}{0.39\textwidth}
		\vspace{-20pt}\raggedright
		\caption{Six successive transmission spectra (frames) for the \unit[16]{mm} waveguide cavity. The large overlap makes it possible to keep track of the slowly varying initial frequency. Inset: the laser diode temperature (Celsius) at which each frame was taken. The arrow indicates the order in which the frames were taken.}
		\label{fig:UncalibratedFramesExample}
	\end{minipage}
\end{figure}
	The \unit[500]{GHz} range of interest exceeds the \unit[100]{GHz} mode-hop free current-tuning range of the laser. We therefore change the set point of the laser temperature controller, and while the temperature and frequency slowly move towards their new equilibrium (over several minutes), we scan the current rapidly (at \unit[20]{Hz}) over a few free spectral ranges of the waveguide cavity being investigated. This generates a collection of \emph{frames} (spectra) for the transmitted intensity through both the waveguide cavity and the reference cavity. Because the FSR of the latter is accurately known, it is possible to normalise the frequency, i.e. to convert the time axis of each frame to optical frequency by mapping the position of the resonances in the reference cavity spectrum.

\begin{figure}[bt]
	\centering
	\includegraphics[width=\textwidth]{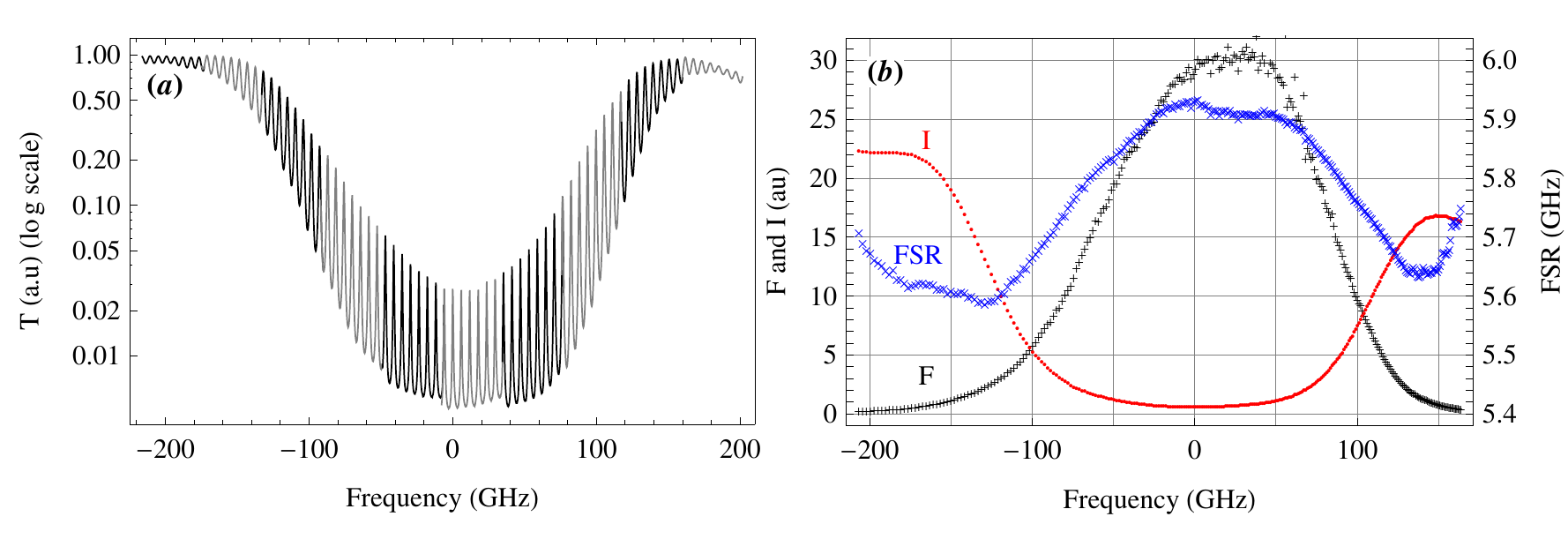}
	\caption{Data taken using a \unit[16]{mm} long waveguide cavity with \unit[1.7]{mm}-long gratings. (a) Log of intensity transmitted by cavity, reconstituted from 13 frames (out of 400), with gray/black parts denoting different frames. In the wings the transmission is close to 100\%, at the centre it oscillates between 0.5\% and 3\%. (b) Black pluses show the coefficient of finesse $F$, blue crosses plot the free spectral range FSR, and red dots indicate the maximum transmitted intensity.}
	\label{fig:CavityParamsExample}
\end{figure}

	Figure \ref{fig:UncalibratedFramesExample} shows a representative sequence of waveguide cavity spectra, exhibiting a large overlap in frequency between consecutive frames as the temperature-induced frequency shift between two frames is smaller than a free spectral range of the waveguide cavity. This means that the initial frame phase $\psi$, as determined by fitting the frequency-normalised spectra with a modified Fabry-Perot Airy function \cite{Born&Wolf}
\begin{equation}
	\frac{I}{1+F \sin ^2 \left[\pi \left(\frac{\nu}{\mathrm{FSR}}+\psi \right) \right]}
	\label{eq:AiryFn}
\end{equation}
(where $F$ is the coefficient of finesse, to be defined later) is sufficiently slowly varying that the unwrapped phase $\psi'$ can be calculated; the quantity FSR$\times \psi'$ is then the starting frequency of the frame. Thus all the frames can be ``stitched'' together to reconstitute a continuous cavity spectrum, as shown in Fig.~\ref{fig:CavityParamsExample}(a) for 13 frames measured with a \unit[16]{mm} cavity. In fitting Eq.~(\ref{eq:AiryFn}) to the data, we allow the free spectral range FSR, the finesse coefficient $F$ and and the amplitude $I$ to vary linearly (or quadratically if necessary) across the frame (e.g. $\mathrm{FSR}(\nu)=\mathrm{FSR}_0+\mathrm{FSR}_1\times\nu+\mathrm{FSR}_2\times\nu^2)$ as they cannot usually be considered constant over the frequency span of a given frame. The three series of data points in Fig.~\ref{fig:CavityParamsExample}(b) show the starting values of these parameters ($\mathrm{FSR}_0$, $F_0$ and $I_0$) for 400 frames that were measured. On the high-frequency side of the band, $F$ drops faster and the transmitted intensity rises more slowly than on the low-frequency side. This asymmetry may be explained by stronger scattering loss and/or by the appearance of cladding modes at higher frequency. These imperfections are not included in our coupled mode theory.

In the following, we devote our attention to the FSR. In order to improve the signal-to-noise ratio, a least-squares fitting method was used to determine the best FSR value for each distinct resonance order in the cavity spectrum. When a resonance peak of order $p$ appears in $q$ consecutive frames, we combine the $\mathrm{FSR}_0$ values associated with these frames to form a single averaged value.

\subsection{Group delay as a function of frequency}\label{Lg_profile}

\begin{figure}[b!]
	\centering
	\includegraphics[width=\textwidth]{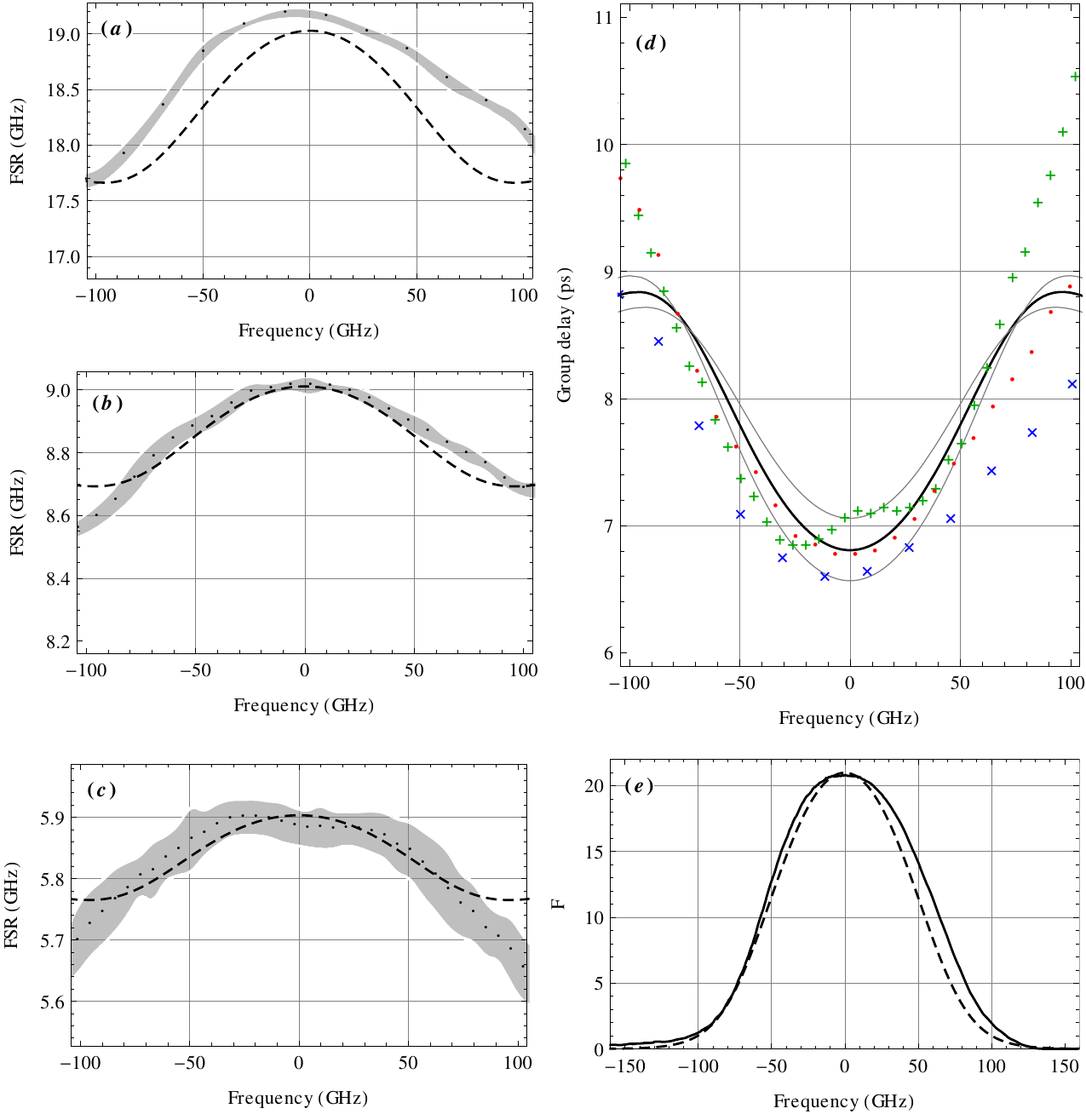}
	\caption{{\bf (a)-(c)}: free spectral range versus frequency for cavities of spacing (a) \unit[16]{mm}, (b) \unit[10]{mm}, and (c) \unit[4]{mm}. Points: experimental data with uncertainties shown as shaded area. Dashed line: theoretical FSR derived from coupled mode theory. {\bf (d)} Grating group delay $\tau_g$. Blue crosses, red dots and green pluses are derived from the data in (a), (b) and (c) respectively, while the solid black line is calculated from coupled mode theory fitted to the data with index contrast as the only fit parameter. {\bf (e)}: Measured (solid) and theoretical (dashed line) finesse of the cavity in (b), fitted by adjusting the propagation loss and the index contrast. }
	\label{fig:Lg_experiment}
\end{figure}	

	In this way, we have analysed the transmission spectra of three grating cavities with spacings $L=4$, 10 and \unit[16]{mm} between the front faces of the Bragg mirrors. Each mirror is \unit[1.7]{mm} long, corresponding to approximately 6400 grating planes. The variation of the FSR with frequency for each cavity is shown by the dots in Figures \ref{fig:Lg_experiment} (a), (b) and (c). As expected the FSR reaches a maximum at the central wavelength of the gratings, where the reflectivity is highest, and therefore the field penetration $L_g$ into the gratings is smallest.

	We take the frequency at which the FSR peaks to be the Bragg frequency $f_B$. Averaged over five such cavities we obtain the value $f_B=\unit[3.8422(1)\times 10^{14}]{Hz}$. Together with the known grating spacing $\Lambda_B=\unit[267.3]{nm}$, this yields a mean refractive index $n=1.4595(1)$ through the relation $n=\frac{c}{2 f_B \Lambda_B}$. The group delay $\tau_g=\frac{n}{c}(L_{\mathrm{eff}}-L)$ is then given by

\begin{equation}
	\tau_g = \frac{1}{2\mathrm{FSR}}-\frac{n L}{c}
	\label{}
\end{equation}
and is plotted in Figure \ref{fig:Lg_experiment}(d) for all three cavities. Near the centre of the band, there is good agreement between the results from the three cavities and with the theory, represented by the solid line, with $\Delta n = 3.9(5)\times 10^{-4}$ being the only free parameter. The two thin gray lines are the theoretical group delay with $\Delta n = 3.85$ and $3.95\times 10^{-4}$ and represent the theoretical error bar; they encompass all the data points near the Bragg frequency. Away from that frequency, however, there are clear systematic differences between the three data sets. These could be due to variation from one grating to another in the apodisation profile. The theory also differs significantly from all three data sets in that it predicts a much faster and more pronounced flattening off of the group delay away from the Bragg frequency. This seems to suggest that the real apodisation profile is not in fact Gaussian, but we do not have enough information to reconstruct what it is.

 

	Finally, we turn to Figure \ref{fig:Lg_experiment}(e) which shows the measured finesse coefficient $F$ for the \unit[10]{mm}-long cavity, plotted against frequency (solid line). For a theoretical cavity with grating reflectivity $R$ and single-pass attenuation $\gamma=\exp(-\alpha L_{\mathrm{eff}})$ due to propagation loss, one finds that $F=\frac{4R \gamma}{\left(1-R \gamma\right)^2}$.  By adjusting $\gamma$ and changing $R$ through the choice of $\Delta n$, we obtain a theoretical curve (dashed) that corresponds reasonably well with the measurements when $\Delta n=5.5\times10^{-4}$ and the absorption coefficient is $\alpha=\unit[0.9]{dB/cm}$. It is difficult to assign error bars here because, once again, there is a systematic difference between theory and experiment in the wings of the band, making this method of measuring $\Delta n$ inferior to the group delay method described above. The primary purpose of this measurement is to estimate the absorption coefficient $\alpha$ which we consider to be accurate within about 20\%.



It is instructive to compare this $\alpha=\unit[0.9(2)]{dB/cm}$ with the $\unit[0.235(6)]{dB/cm}$ loss measured in nearly identical waveguides \cite{Rogers2010} at \unit[1550]{nm}. This indicates that only a small part or the loss at 1550\,nm, 20\% or less, can be due to Rayleigh scattering, which scales as $\lambda^{-4}$. We believe that the total loss may be significantly reduced in future by improving control over the planar silica layers.

\subsection{Group delay as a function of mirror length}
\begin{figure}[t]
	\centering
	\includegraphics[width=\textwidth]{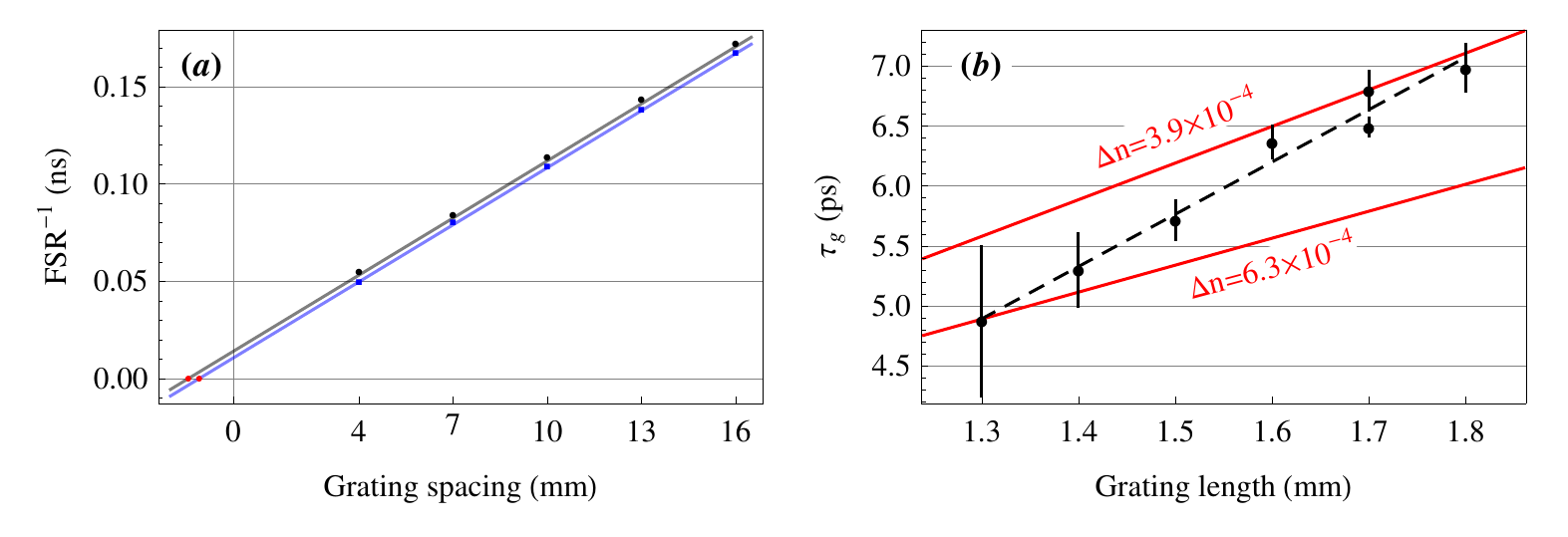}
	\caption{{\bf (a)} $\mathrm{FSR}^{-1}$ plotted versus length $L$ for five cavities formed between \unit[1.4]{mm}-long gratings and five using \unit[1.8]{mm}-long gratings. Determination of group delay by extrapolating  to $\mathrm{FSR}^{-1}=0$, indicated by red dots in the bottom left corner. {\bf (b)} Evolution of group delay as the length of the grating increases. For the penultimate grating, two points are shown: one is derived from the three cavities of Figure \ref{fig:Lg_experiment}, while the second one uses a different set of data including all five cavities. The dashed line is a linear fit to the data. Red lines: coupled mode theory fro two values of index contrast. }
	\label{fig:Lg_All}
\end{figure}
In a second experiment, we used a set of grating cavities designed to study the group delay as a function of the physical length of the grating. The gratings, of six different lengths ranging from 1.3 to \unit[1.8]{mm}, were apodised with a Gaussian whose standard deviation is one quarter of the grating length. For each grating length we made five cavities with mirror spacings $L$ ranging from 4 to \unit[16]{mm}. We then measured the maximum FSR for each cavity, corresponding to the minimum group delay, which occurs at the centre of the stopband. Since the FSR is inversely proportional to the cavity length $L_{\mathrm{eff}}$, a plot of FSR$^{-1}$ versus $L$ yields an intercept on the abscissa of $-2 L_g$, which gives $\tau_g$. Figure~\ref{fig:Lg_All}(a) shows the plots for two series of cavities, one with gratings \unit[1.4]{mm} long and the other \unit[1.8]{mm}. The six values of $\tau_g$ derived from these and the other four sets of cavities are plotted in Fig.~\ref{fig:Lg_All}(b) as black dots with error bars. As indicated by the dashed line, $\tau_g$ increases linearly with the physical length of the grating. In part, this reflects the fact that, because of the apodising function, the field must penetrate more deeply into longer gratings before the index contrast is high enough to produce strong reflection. The upper solid red line in Fig.~\ref{fig:Lg_All}(b) show the group delay given by coupled mode theory for an index contrast of $3.9\times10^{-4}$, the contrast that we already measured in section \ref{Lg_profile} using the cavities formed from long gratings. By contrast, the results obtained using shorter gratings correspond to $\Delta n$ as large as $6.3\times10^{-4}$, the lower red line. We believe that the decrease of $\Delta n$ with grating length is in fact a decrease of index contrast with the time of writing, the gratings having been written in order of increasing length. The waveguide chip is loaded with hydrogen prior to writing in order to increase the photosensitivity, and it is known that enough hydrogen outgasses over the several hours of writing to produce an effect of this size.

\section{Summary and conclusion}
	Cavities integrated into optical waveguides chips are of great interest for chemical and biological analysis on a chip and for applications in quantum optics and quantum information processing. Visible and very near infrared wavelengths are particularly important for these applicaitons.  We are currently developing such cavities incorporating Bragg reflectors, fabricated by writing in a photosensitive doped silica layer using UV laser light. In the course of characterising them, we have measured transmission spectra from which we derive values for the free spectral range as a function of frequency. We have shown how this provides a simple way to measure the delay of light reflected from the mirrors. Using a coupled mode approach, we have used this delay to determine the index contrast of the gratings. These quantities, delay and index contrast, are of sufficiently general interest that our very simple method may be of use in characterising other devices such as fibre Bragg gratings and delay lines, which are widely used in the telecommunications and astronomy.

\section*{Acknowledgements}
This work was supported by the UK's EPSRC, by the Royal Society, and by the EC through the HIP and ACUTE networks.



\begin{thebibliography}{99}
\newcommand{\enquote}[1]{``#1''}

\bibitem{Sparrow2005}
I.~J.~G. Sparrow, G.~D. Emmerson, C.~B.~E. Gawith, and P.~G.~R. Smith,
  \enquote{Planar waveguide hygrometer and state sensor demonstrating supercooled water recognition,} Sensors and Actuators B: Chemical \textbf{107}, 856--860 (2005).

  \bibitem{Hale1973}
G.~M. Hale and M.~R. Querry, \enquote{Optical constants of water in the 200-nm to 200-µm wavelength region,} Appl. Opt. \textbf{12}, 555--563 (1973).

\bibitem{Lepert2011}
G. Lepert, M. Trupke, M.~J. Hartman, M.~B. Plenio, and E.~A. Hinds,
\enquote{Arrays of waveguide-coupled optical cavities that interact strongly with atoms,} arXiv:1109.0886 [quant-ph] (2011).

\bibitem{Svalgaard1994}
M.~Svalgaard, C.~V. Poulsen, A.~Bjarklev, and O.~Poulsen, \enquote{Direct UV writing of buried single-mode channel waveguides in Ge-doped silica films,} Electronics Letters \textbf{30}, 1401--1403 (1994).

\bibitem{Kundys2009}
D.~O. Kundys, J.~C. Gates, S.~Dasgupta, C.~Gawith, and P.~Smith, \enquote{Use of cross-couplers to decrease size of UV written photonic circuits,} IEEE  Photonics Technology Letters \textbf{21}, 947--949 (2009).

\bibitem{Erdogan1997}
T.~Erdogan, \enquote{Fiber grating spectra,} Journal of Lightwave Technology \textbf{15}, 1277--1294 (1997).

\bibitem{Barmenkov2006}
Y.~O. Barmenkov, D.~Zalvidea, S.~Torres-Peir{\'{o}}, J.~L. Cruz, and M.~V. Andr{\'{e}}s, \enquote{Effective length of short Fabry-Perot cavity formed by uniform fiber Bragg gratings,} Opt. Express \textbf{14}, 6394--6399 (2006).

\bibitem{Rogers2010}
H.~L. Rogers, S.~Ambran, C.~Holmes, P.~G.~R. Smith, and J.~C. Gates,
  \enquote{In situ loss measurement of direct UV-written waveguides using integrated Bragg gratings,} Opt. Lett. \textbf{35}, 2849--2851 (2010).

\bibitem{Born&Wolf}
M. Born and E. Wolf, \emph{Principles of Optics}, Cambridge University Press (1999)


\end{thebibliography}
\end{document}